\documentclass[iop,apj]{emulateapj}

\usepackage{apjfonts}
\usepackage{psfig}
\usepackage[squaren,Gray]{SIunits}

\def\aap{A\&A}

\def\apjl{ApJ}
\def\apjs{ApJS}

\def\lir{$L_{\rm IR}$}

\def\l1.4{$L_{\rm 1.4GHz}$} \def\s1.4{$S_{\rm 1.4GHz}$}

\def\Td{$T_{\rm d}$}

\def\gs{\mathrel{\raise0.35ex\hbox{$\scriptstyle >$}\kern-0.6em
\lower0.40ex\hbox{{$\scriptstyle \sim$}}}}
\def\ls{\mathrel{\raise0.35ex\hbox{$\scriptstyle <$}\kern-0.6em
\lower0.40ex\hbox{{$\scriptstyle \sim$}}}}
\def\m@th{\mathsurround=0pt }
\def\eqalign#1{\null\,\vcenter{\openup1\jot \m@th
 \ialign{\strut\hfil$\displaystyle{##}$&$\displaystyle{{}##}$\hfil
 \crcr#1\crcr}}\,}

\begin{document}

\title{IMAGING THE ENVIRONMENT OF A {\textit z} = 6.3
  SUBMILLIMETER GALAXY WITH SCUBA-2}

\shorttitle{SCUBA-2 imaging of an SMG at {\it z} = 6.3}
\shortauthors{Robson et al.}

\author{E.\,I.~Robson\altaffilmark{1}}
\author{R.\,J.~Ivison\altaffilmark{2,3}}
\author{Ian~Smail\altaffilmark{4}}
\author{W.\,S.~Holland\altaffilmark{1,3}}
\author{J.\,E.~Geach\altaffilmark{5}}
\author{A.\,G.~Gibb\altaffilmark{6}}
\author{D.~Riechers\altaffilmark{7}}
\author{P.\,A.\,R.~Ade\altaffilmark{8}}
\author{D.~Bintley\altaffilmark{9}}
\author{J.~Bock\altaffilmark{10}}
\author{E.\,L.~Chapin\altaffilmark{11}}
\author{S.\,C.~Chapman\altaffilmark{12}}
\author{D.\,L.~Clements\altaffilmark{13}}
\author{A.~Conley\altaffilmark{14}}
\author{A.~Cooray\altaffilmark{15}}
\author{J.\,S.~Dunlop\altaffilmark{3}}
\author{D.~Farrah\altaffilmark{16}}
\author{M.~Fich\altaffilmark{17}}
\author{Hai~Fu\altaffilmark{15}}
\author{T.~Jenness\altaffilmark{7}}
\author{N.~Laporte\altaffilmark{18}}
\author{S.\,J.~Oliver\altaffilmark{19}}
\author{A.~Omont\altaffilmark{20,21}}
\author{I.~P\'erez-Fournon\altaffilmark{18,22}}
\author{Douglas~Scott\altaffilmark{6}}
\author{A.\,M.~Swinbank\altaffilmark{4}}
\author{J.~Wardlow\altaffilmark{23}}

\altaffiltext{1}{UK ATC, Royal Observatory, Blackford Hill, Edinburgh EH9 3HJ, UK}
\altaffiltext{2}{ESO, Karl Schwarzschild Strasse 2, D-85748 Garching, Germany}
\altaffiltext{3}{Institute for Astronomy, University of Edinburgh,
  Royal Observatory, Blackford Hill, Edinburgh EH9 3HJ, UK}
\altaffiltext{4}{Institute for Computational Cosmology, Durham
  University, South Road, Durham DH1 3LE, UK}
\altaffiltext{5}{Centre for Astrophysics Research, University of
  Hertfordshire, Hatfield AL10 9AB, UK}
\altaffiltext{6}{Dept of Physics \& Astronomy, University of British Columbia,
  6224 Agricultural Road, Vancouver, BC V6T 1Z1, Canada}
\altaffiltext{7}{Astronomy Department, Cornell University, Ithaca, NY 14853}
\altaffiltext{8}{Astronomy and Instrumentation Group, Cardiff
  University, Cardiff, Wales, UK}
\altaffiltext{9}{Joint Astronomy Centre, 660 N.\ Ah\-ok\-u Place,
  University Park, Hilo, Hawaii 96720}
\altaffiltext{10}{Jet Propulsion Laboratory, National Aeronautics and
  Space Administration, Pasadena, CA 91109}
\altaffiltext{11}{XMM SOC, ESAC, Apartado 79, 28691 Villaneueva de la
  Canada, Madrid, Spain}
\altaffiltext{12}{Dalhousie University, Department of Physics and Atmospheric Science, Coburg Road, Halifax B3H 1A6, Canada}
\altaffiltext{13}{Astrophysics Group, Imperial College London,
  Blackett Laboratory, Prince Consort Road, London SW7 2AZ, UK}
\altaffiltext{14}{Center for Astrophysics and Space Astronomy, 389
  UCB, University of Colorado, Boulder, CO 80309}
\altaffiltext{15}{Department of Physics and Astronomy, University of
  California, Irvine, CA 92697}
\altaffiltext{16}{Department of Physics, Virginia Tech, Blacksburg, VA 24061}
\altaffiltext{17}{Department of Physics and Astronomy, University of Waterloo,
  Waterloo, Ontario N2L 3G1, Canada}
\altaffiltext{18}{IAC, E-38200 La  Laguna, Tenerife, Spain}
\altaffiltext{19}{Astronomy Centre, Department of Physics and
  Astronomy, University of Sussex, Brighton, BN1 9QH, UK}
\altaffiltext{20}{UPMC Univ Paris 06, UMR 7095, IAP, 75014, Paris, France}
\altaffiltext{21}{CNRS, UMR7095, IAP, F-75014, Paris, France}
\altaffiltext{22}{Departamento de Astrofisica, Universidad de La
  Laguna, E-38205 La Laguna, Tenerife, Spain}
\altaffiltext{23}{Dark Cosmology Centre, Niels Bohr Institute,
  University of Copenhagen, Denmark}

\slugcomment{To be submitted to The Astrophysical Journal}

\begin{abstract}
  We describe a search for submillimeter emission in the vicinity of
  one of the most distant, luminous galaxies known, HerMES FLS3
  at $z=6.34$, exploiting it as a signpost to a potentially biased
  region of the early Universe, as might be expected in hierarchical
  structure formation models. Imaging to the confusion limit with the
  innovative, wide-field submillimeter bolometer camera, SCUBA-2, we
  are sensitive to colder and/or less luminous galaxies in the
  surroundings of HFLS3. We use the Millennium Simulation to
  illustrate that HFLS3 may be expected to have companions if it is as
  massive as claimed, but find no significant evidence from the
  surface density of SCUBA-2 galaxies in its vicinity, or their
  colors, that HFLS3 marks an over-density of dusty, star-forming
  galaxies. We cannot rule out the presence of dusty neighbours with
  confidence, but deeper 450-$\mu$m imaging has the potential to more
  tightly constrain the redshifts of nearby galaxies, at least one of
  which likely lies at $z\gs5$. If associations with HFLS3 can be
  ruled out, this could be taken as evidence that HFLS3 is less biased
  than a simple extrapolation of the Millennium Simulation may
  imply. This could suggest either that it represents a rare
  short-lived, but highly luminous, phase in the evolution of an
  otherwise typical galaxy, or that this system has suffered
  amplification due to a foreground gravitational lens and so is not
  as intrinsically luminous as claimed.
\end{abstract}

\keywords{galaxies: high-redshift --- galaxies: starburst ---
  submillimeter: galaxies --- infrared: galaxies --- radio continuum:
  galaxies}

\section{Introduction}
\label{intro}

%
% Figure 1
%
\begin{figure*}
\centerline{\psfig{file=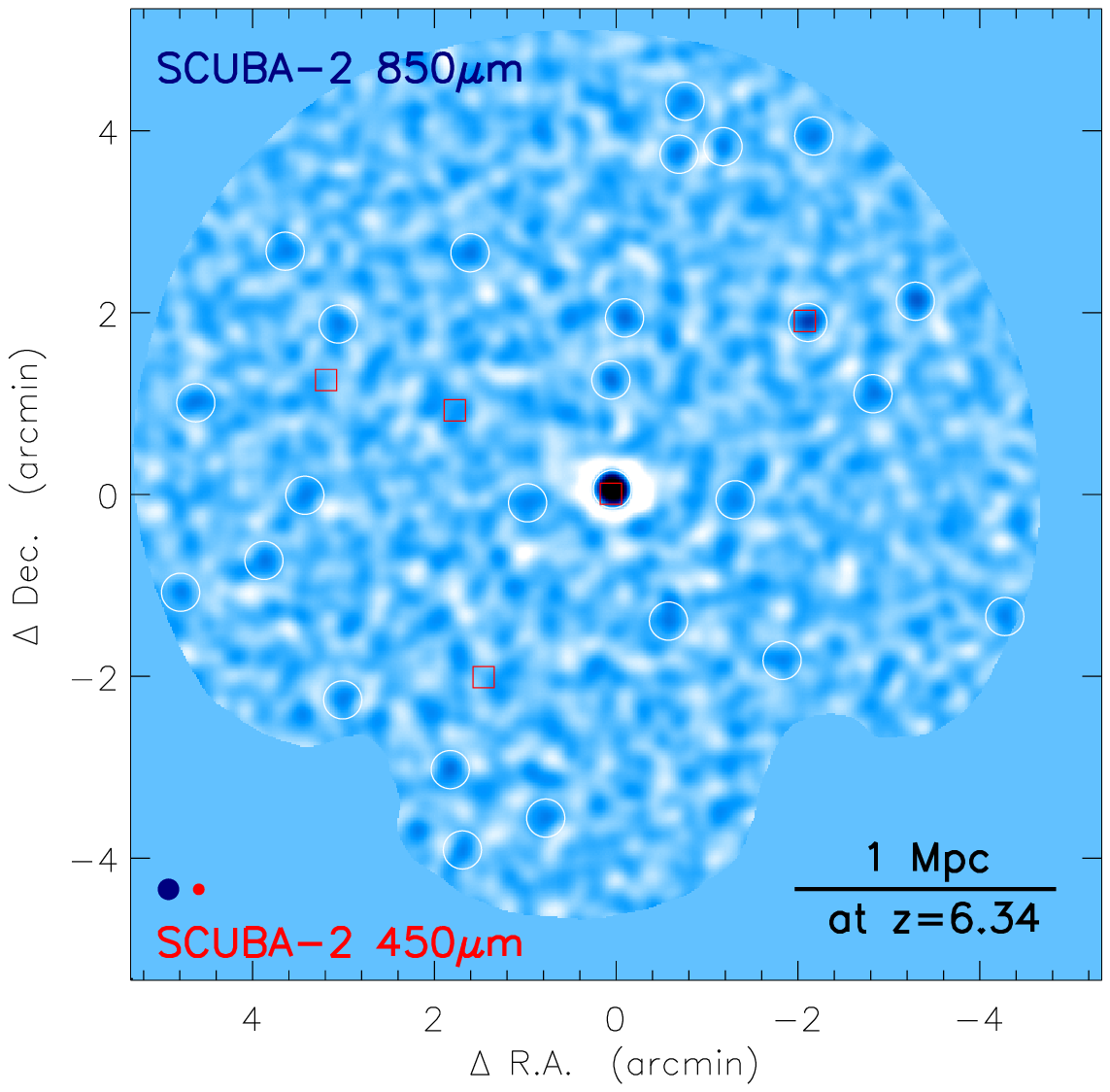,width=3.4in,angle=0}
\psfig{file=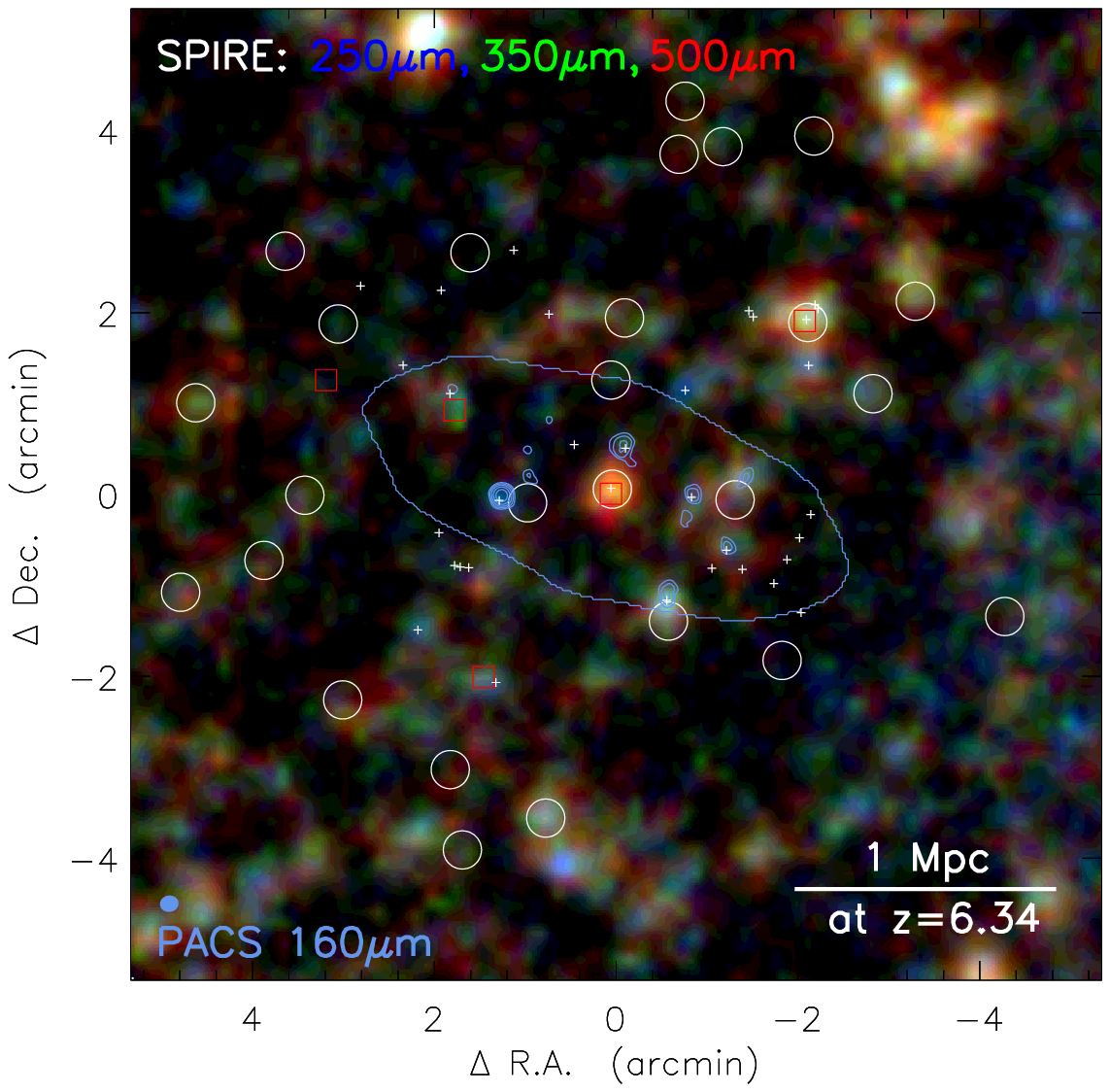,width=3.4in,angle=0}}
\caption{{\it Left:} SCUBA-2 imaging at 850\,$\mu$m, with 450- and
  850-$\mu$m sources marked by red squares and white circles,
  respectively, for the 67.2\,arcmin$^2$ where $\sigma_{850}\le
  1.5$\,mJy\,beam$^{-1}$. The negative bowl around HFLS3 is a typical
  artifact of the filtering procedures employed here. The FWHM of the
  SCUBA-2 beams are shown as solid ellipses. {\it Right:} Three-color
  representation of the data obtained using SPIRE at 250, 350 and
  500\,$\mu$m for the same field around HFLS3, superimposed with blue
  PACS 160-$\mu$m contours.  Several of the SCUBA-2 850-$\mu$m sources
  are associated with green SPIRE sources -- those with SEDs peaking
  at 350\,$\mu$m, consistent with $z\approx 2$; others have no obvious
  SPIRE counterparts and may lie at considerably higher redshifts. The
  region over which PACS sensitivity is better than half the best is
  outlined in blue. Positions of faint 1.4-GHz sources from the
  $\sim1.3''$-resolution, $\sigma\sim11\,\mu$Jy\,beam$^{-1}$ Karl G.\
  Jansky Very Large Array imaging described in \citet{riechers13} are
  marked `+' (the radio catalog covers only $\approx$25\% of the
  region shown, hence the detection rate is unremarkable). N is up; E
  is left; offsets from $\alpha_{2000}=17$:06:47.8,
  $\delta_{2000}=+58$:46:23 are marked. }
\label{fig:imaging}
\end{figure*} 

Dust extinction and a profusion of less luminous foreground galaxies
makes it difficult to select high-redshift ultraluminous star-forming
galaxies ($L_{\rm IR}\geq 10^{12}$\,L$_\odot$) at rest-frame
ultraviolet/optical wavelengths.  Although extinction is not an issue
at radio wavelengths, an unfavourable $K$-correction works against
detecting the highest redshift examples, $z\gg 3$. Since the advent of
large-format submillimeter (submm) cameras such as the Submillimeter
Common-User Bolometer Array \citep[SCUBA --][]{holland99}, however, it
has been possible to exploit the {\it negative} $K$-correction in the
submm waveband to select dusty, star-forming galaxies (submm-selected
galaxies, or SMGs) almost independently of their redshift
\citep[e.g.][]{fran91, blainlongair93}.

The scope of this field has been substantially expanded by {\it
  Herschel} \citep{pilbratt10} which has surveyed approximately a
hundred square degrees of extragalactic sky to the confusion limit at
500\,$\mu$m \citep[as defined by][]{nguyen10}, with simultaneous
imaging at 250 and 350\,$\mu$m, using the SPIRE instrument
\citep{griffin10}. A SPIRE image of the {\it Spitzer} First Look
Survey (FLS) field, obtained as part of the {\it Herschel}
Multi-Tiered Extragalactic Survey
\citep[HerMES\footnote{hermes.sussex.ac.uk} --][]{oliver12}, led to
the discovery of 1HERMES\,S350\,J170647.8+584623 \citep[hereafter
HFLS3 --][]{riechers13, dowell14} as an unusually red SPIRE source
with $S_{250}<S_{350}<S_{500}$, i.e.\ with its thermal dust peak
within or beyond the 500-$\mu$m band \citep[see also][]{cox11,
  combes12, rawle14}. Some of these ``500-$\mu$m risers'' are in fact
due to synchrotron emission from bright, flat-spectrum radio quasars
\citep[e.g.][]{jenness10}, but HFLS3 does not exhibit such powerful
AGN-driven radio emission.  Panchromatic spectral-line observations
place HFLS3 at $z=6.34$ via the detection of H$_2$O, CO, OH, OH$^+$,
NH$_3$, [C\,{\sc i}] and [C\,{\sc ii}] emission and absorption
lines. Its continuum spectral energy distribution (SED) is consistent
with a characteristic dust temperature, \Td\ = 56\,\kelvin, and a dust
mass of $1.1\times 10^9$\,M$_\odot$. Its infrared luminosity, \lir\ =
$2.9\times 10^{13}$\,L$_\odot$, suggests a star-formation rate (SFR)
of $2900\,\mu_{\rm L}^{-1}$\,M$_\odot$\,yr$^{-1}$ for a
\citet{chabrier03} initial mass function, where the lensing
magnification suffered by HFLS3 due to a foreground galaxy less than
2$''$ away has been estimated to be in the range $\mu_{\rm
  L}=1.2$--1.5 \citep[][although \citealt{cooray14} estimate $\mu_{\rm
  L}=2.2\pm 0.3$]{riechers13}.

It is expected that the most massive galaxies found at very high
redshifts grew in (and thus signpost) the densest peaks in the early
Universe, making them useful tracers of distant proto-clusters. Above
$z \sim 6$, such sources may also contribute to the rapid evolution of
the neutral fraction of the Universe, during the so-called `era of
reionisation', and to the earliest phase of enrichment of the
interstellar medium in galaxies, less than 1\,Gyr after the Big
Bang. They may also host the highest redshift quasars.  In the submm
regime, to explore distant galaxies and their environments we have
observed radio galaxies and quasars, typically detecting factor
$\sim2$--4$\times$ over-densities of submm companions around these
signposts \citep[e.g.][]{ivison00, stevens03, stevens10, robson04,
  priddey08}. Here, we continue this tradition, targeting the most
distant known submm galaxy, HFLS3 at $z= 6.34$, with the 10,000-pixel
SCUBA-2 bolometer camera \citep{holland13}, which is more sensitive
than {\it Herschel} to cold dust in high-redshift galaxies.

In \S\ref{observations} we describe our SCUBA-2 observations of the
field surrounding HFLS3, after its discovery with SPIRE aboard {\it
  Herschel}, and our reduction of those data. In \S\ref{results} we
analyze the surface density of SCUBA-2 galaxies in the field, and
their color, and discuss whether there is any evidence that HFLS3
inhabits an over-dense region of the Universe, as might be expected in
hierarchical structure-formation models \citep[e.g.][]{kaufman99,
  springel05}. We finish with our conclusions in
\S\ref{conclusions}. Throughout, we adopt a cosmology with $H_0 =
71$\,km\,s$^{-1}$\,Mpc$^{-1}$, $\Omega_{\rm m}=0.27$ and
$\Omega_\Lambda = 0.73$, so 1$''$ equates to 5.7\,kpc at $z=6.34$.

\section{Observations and data reduction}
\label{observations}

\subsection{SCUBA-2 imaging and catalogues}
\label{scuba2}

Data were obtained simultaneously at 450 and 850\,$\mu$m in 2011
September 23-24 and 2012 March 14 (project M11BGT01) using SCUBA-2 on
the 15-m James Clerk Maxwell Telescope (JCMT). The observations were
taken with the constant speed {\sc daisy} pattern, which provides
uniform exposure-time coverage in the central 3$^\prime$-diameter
region of a field, and useful coverage over 12$^\prime$. A total of
6.9\,hr was spent integrating on HFLS3. Observing conditions were good
or excellent, with precipitable-water-vapor levels typically 1\,mm or
less, corresponding to a 225-GHz optical depth of 0.05. The data were
calibrated in flux density against the primary calibrators Uranus and
Mars, and also secondary sources CRL\,618 and CRL\,2688 from the JCMT
calibrator list \citep{dempsey13}, with estimated calibration
uncertainties amounting to 5\% at 850\,$\mu$m and 10\% at 450\,$\mu$m.

The data were reduced using the Dynamic Iterative Map-Maker within the
{\sc starlink smurf} package \citep{chapin13} called from the {\sc
  orac-dr} automated pipeline \citep{cavanagh08}. The chosen recipe
accounted for attenuation of the signal as a result of time-series
filtering and removed residual low-frequency noise from the map using
a jack-knife method. The maps were made using inverse-variance
weighting, with 1$^{\prime\prime}$ pixels at both wavelengths, before
application of a matched filter \citep[e.g.][]{chapin11}.

The map-maker used a `blank field' configuration file, optimized for
faint, unresolved or compact sources. This applies a high-pass filter
with a spatial cutoff of 200$^{\prime\prime}$, corresponding to about
0.8\,Hz for a typical scanning speed of
155$^{\prime\prime}$\,s$^{-1}$. This removes the majority of
low-frequency (large spatial scale) noise, while the remainder is
removed using a Fourier-space whitening filter. This is derived from
the power spectrum of the central 9$^\prime$ region of a jack-knife
map, produced from two independent halves of the total dataset.

This filtering attenuates the peak signal of sources in the map. To
estimate the magnitude of this effect, the pipeline re-makes each map
with a fake 10-Jy Gaussian added to the raw data, offset from the
nominal map centre by 30$^{\prime\prime}$ to avoid contamination by
any target at the map centre. The amplitude of the fake Gaussian in
the output map is measured to determine a correction factor. The
standard flux conversion factor (FCF), as determined from observations
of primary and secondary calibrators, is then multiplied by this
factor (1.17 and 1.15 at 850 and 450\,$\mu$m, respectively) and
applied to the final image to give a map calibrated in
Jy\,beam$^{-1}$. The maps with the fake Gaussian are also used to form
the point spread function (PSF) for the matched filter since they
reflect the effective point-source transfer function of the map-maker.

The SCUBA-2 850- and 450-$\mu$m images shown in Fig.~\ref{fig:imaging}
reach noise levels of 0.9 and 5.0\,mJy\,beam$^{-1}$ over the central
3$^\prime$-diameter regions, yielding detections of HFLS3 at
approximately the 41- and 7-$\sigma$ levels, respectively. At
850\,$\mu$m, the central 67.2\,arcmin$^2$ of the map has a noise level
of 1.5\,mJy\,beam$^{-1}$ or better.  The astrometry of the SCUBA-2
images was found to be accurate to better than 1$^{\prime\prime}$ by
stacking at the positions of 3.6-$\mu$m and 1.4-GHz sources in the
field.

Following \citet{geach13}, we create a catalogue of sources from the
450- and 850-$\mu$m images by searching for peaks in the
beam-convolved signal-to-noise ratio maps, recording their
coordinates, flux densities and local noise levels. We then mask a
region $1.5\times$ the beam size and then repeat the search. Above a
signal-to-noise level of 3.75 the contamination rate due to false
detections is below 5\%. We adopt this as our detection threshold,
listing the 26 sources with 850-$\mu$m flux density uncertainties
below 1.5\,mJy in Table~\ref{tab:sample}, alongside 450-$\mu$m sources
selected from the same area at the same significance threshold.

We calculate our completeness levels and flux boosting following
\citet{geach13}, who followed \citet{weiss09}, injecting $10^5$
artificial point sources into a map with the same noise properties as
the real image.  We correct for false positives using the jack-knife
map.

\begin{deluxetable}{lcc}
\tabletypesize{\scriptsize}
\tablecaption{Sources detected at 850 and 450\,$\mu$m near
  HFLS3.\label{tab:sample}}
\tablewidth{0pt}
\tablehead{
\colhead{IAU name} &
\colhead{$S$ /mJy\tablenotemark{a}} & 
\colhead{S/N}}
\startdata
850\,$\mu$m:&&\\ 
S2FLS850\,J170647.67+584623.0\tablenotemark{b} & $35.4\pm 0.9$ & 40.9\\
S2FLS850\,J170631.07+584812.9 & $11.4\pm 1.3$ & 8.6 \\
S2FLS850\,J170621.93+584826.8 & $9.3\pm  1.4$ & 6.5 \\
S2FLS850\,J170646.64+584816.0 & $5.6\pm  1.0$ & 5.5 \\
S2FLS850\,J170647.80+584735.0 & $4.9\pm  0.9$ & 5.3 \\
S2FLS850\,J170701.41+584318.0 & $7.5\pm  1.4$ & 5.3 \\
S2FLS850\,J170625.54+584725.9 & $6.3\pm  1.3$ & 4.8 \\
S2FLS850\,J170642.92+584456.0 & $4.7\pm  1.0$ & 4.6 \\
S2FLS850\,J170717.24+584535.8 & $5.8\pm  1.3$ & 4.6 \\
S2FLS850\,J170659.77+584859.0 & $5.6\pm  1.2$ & 4.6 \\
S2FLS850\,J170723.05+584719.7 & $6.1\pm  1.4$ & 4.5 \\
S2FLS850\,J170642.00+585004.0 & $5.4\pm  1.2$ & 4.5 \\
S2FLS850\,J170630.54+585015.9 & $6.2\pm  1.4$ & 4.4 \\
S2FLS850\,J170710.97+584811.9 & $5.4\pm  1.3$ & 4.2 \\
S2FLS850\,J170653.32+584246.0 & $5.5\pm  1.3$ & 4.2 \\
S2FLS850\,J170713.78+584618.8 & $4.9\pm  1.2$ & 4.2 \\
S2FLS850\,J170638.27+585009.0 & $5.3\pm  1.3$ & 4.2 \\
S2FLS850\,J170710.54+584403.9 & $6.1\pm  1.5$ & 4.1 \\
S2FLS850\,J170724.30+584514.7 & $5.9\pm  1.5$ & 4.0 \\
S2FLS850\,J170700.38+584225.0 & $5.9\pm  1.5$ & 4.0 \\
S2FLS850\,J170715.48+584859.8 & $5.4\pm  1.3$ & 4.0 \\
S2FLS850\,J170614.39+584458.7 & $6.0\pm  1.5$ & 4.0 \\
S2FLS850\,J170641.49+585039.0 & $5.3\pm  1.3$ & 4.0 \\
S2FLS850\,J170637.26+584616.0 & $3.9\pm  1.0$ & 3.9 \\
S2FLS850\,J170633.28+584430.0 & $5.2\pm  1.3$ & 3.8 \\
S2FLS850\,J170654.87+584614.0 & $3.3\pm  0.9$ & 3.8 \\
\noalign{\smallskip}
450\,$\mu$m:&&\\ 
S2FLS450\,J170647.80+584620.0\tablenotemark{b} & $39.8\pm 5.5$ & 7.3\\
S2FLS450\,J170701.05+584715.0 & $29.4\pm 6.9$ & 4.3\\
S2FLS450\,J170631.33+584813.9 & $36.3\pm 8.8$ & 4.1\\
S2FLS450\,J170636.12+584224.0 & $32.5\pm 8.3$ & 3.9\\
S2FLS450\,J170658.59+584419.0 & $30.2\pm 7.8$ & 3.9\\
S2FLS450\,J170711.99+584734.9 & $34.5\pm 8.9$ & 3.9\\
\enddata
\tablenotetext{a}{Deboosted flux densities; errors exclude 5 and 10\%
  calibration uncertainties at 850 and 450\,$\mu$m, respectively.}
\tablenotetext{b}{HFLS3.}
\end{deluxetable}

\subsection{{\it Herschel} imaging}
\label{herschel}

The acquisition and reduction of 16.8\,hrs of {\it Herschel} SPIRE and
(shallow) PACS data for the FLS field (OD159, 164) as part of HerMES
is described in detail by \citet{oliver12}. The SPIRE data, which are
confusion limited, are shown as a three-color image in
Fig.~\ref{fig:imaging}.

We have obtained much deeper data from PACS \citep{poglitsch10} via a
3.9-hr integration as part of programme {\sc ot2\_driecher\_3}
(OD1329) \citep[see][for further details]{riechers13}.  Observations
were carried out on 2013 January 01 in mini-scan mapping mode ($4
\times 15$ repeats), using the 70- plus 160-$\mu$m parallel mode and
the 110- plus 160-$\mu$m parallel mode for one orthogonal cross-scan
pair each. In the 70-, 110-, and 160-$\mu$m bands, the r.m.s.\
sensitivities at the position of HFLS3 are 0.67, 0.73, and
1.35\,mJy\,beam$^{-1}$, respectively. Data reduction and mosaicing
were carried out using standard procedures. The absolute flux density
scale is accurate to 5\%. The 160-$\mu$m PACS image, the only one
potentially useful in the context of faint, distant galaxies, is shown
in Fig.~\ref{fig:imaging}.

The 250-, 350- and 500-$\mu$m flux densities, $S_{250}$, $S_{350}$ and
$S_{500}$, at the positions\footnote{Positions are known to
  $\sigma_{\rm pos}\approx 2.2''$ even for the least significant SMGs,
  a small fraction of the beam-convolved SPIRE point spread function.}
of the 19 SMGs discussed in \S\ref{scuba2} were determined using
beam-convolved SPIRE maps.  None of our SMGs lie near bright SPIRE
sources so we expect the uncertainties associated with these flux
densities should be close to the typical confusion levels, $\approx
6$\,mJy \citep[where hereafter we adopt $\sigma_{\rm conf}$ in each
SPIRE band from][]{nguyen10}.

\section{Results, analysis and discussion}
\label{results}

HFLS3 dominates the submm sky in the 67.2-arcmin$^2$ (8\,Mpc$^2$)
region we have mapped at 850\,$\mu$m with SCUBA-2, being three times
brighter than the next-brightest submm emitter
(Fig.~\ref{fig:imaging}; Table~\ref{tab:sample}). At 450\,$\mu$m,
HFLS3 is the brightest SMG in the region, despite the peak of its SED
having moved beyond that filter; it is one of two sources detected
formally at both 450 and 850\,$\mu$m. Perhaps unsurprisingly, there
are no sources in common between the SCUBA-2 and PACS images.

We see no evidence for an over-density of SMGs on $<$\,1.5-Mpc scales
around the position of HFLS3 in either our 450 or 850-$\mu$m maps
(Fig.~\ref{fig:imaging}).

\subsection{Number counts relative to blank fields}
\label{counts}

%
% Figure 2
%
\begin{figure}
\centerline{\psfig{file=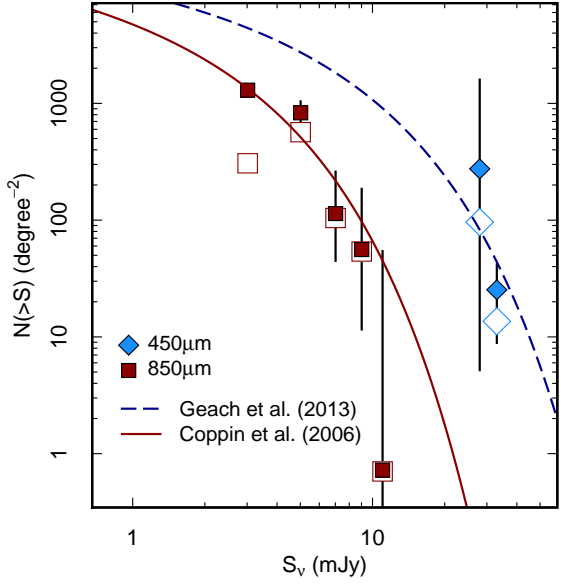,width=3.35in,angle=270}}
\vspace{-2mm} 
\caption{Source counts in several flux density bins at 450 and
  850\,$\mu$m in the 67.2-arcmin$^2$ region around HFLS3, excluding
  HFLS3 itself, relative to those found in typical blank fields by
  \citet{coppin06} and \citet{geach13}.  We see no indication of a
  strong excess of submm emitters across the field, compared to
  blank-field counts, which suggests there is no over-density of SMGs
  on $\sim$\,1.5-Mpc scales around HFLS3. The raw counts, uncorrected
  for incompleteness, are shown as open symbols.}
\label{fig:counts}
\end{figure} 
 
Although no obvious cluster of submm emitters is visible near HFLS3 in
Fig.~\ref{fig:imaging}, we must ask whether the entire 8-Mpc$^2$ field
might be over-dense in SMGs?  Fig.~\ref{fig:counts} shows the density
of sources brighter than $S_\nu$ at 450 and 850\,$\mu$m -- extracted
at the 3.75-$\sigma$ level and corrected for incompleteness using the
analysis discussed in \S\ref{scuba2}, excluding HFLS3 itself --
relative to the source density seen in typical blank-field surveys,
where the same techniques have been used to construct catalogues and
correct for incompleteness \citep{coppin06, geach13}.  The only hint
of an over-density comes in the 850-$\mu$m bin at 5\,mJy, but a
$\approx2$-$\sigma$ deviation is not unusual when plotting seven
points. Simplifying matters, then, by taking only one large bin above
our detection threshold at 850\,$\mu$m, we find 26 sources (25 if we
ignore HFLS3) against an expectation of 20 from \citealt{coppin06} --
on the face of it, a $\approx$1-$\sigma$ over-density. When we take
into account the errors associated with the \citealt{coppin06} counts,
we expect to see 26 (25) sources in a random field of the same size
13\% (19\%) of the time, so the result obtained intuitively from
studying Fig.~\ref{fig:counts} is confirmed -- any excess is
statistically unconvincing.  No significant over-density of SMGs is
apparent in the vicinity of HFLS3, at least not on Mpc scales at flux
densities above the JCMT confusion limit.

\subsection{Redshift constraints for sources in the field}
\label{zconstraints}

%
% Figure 3
%
\begin{figure}
\centerline{\psfig{file=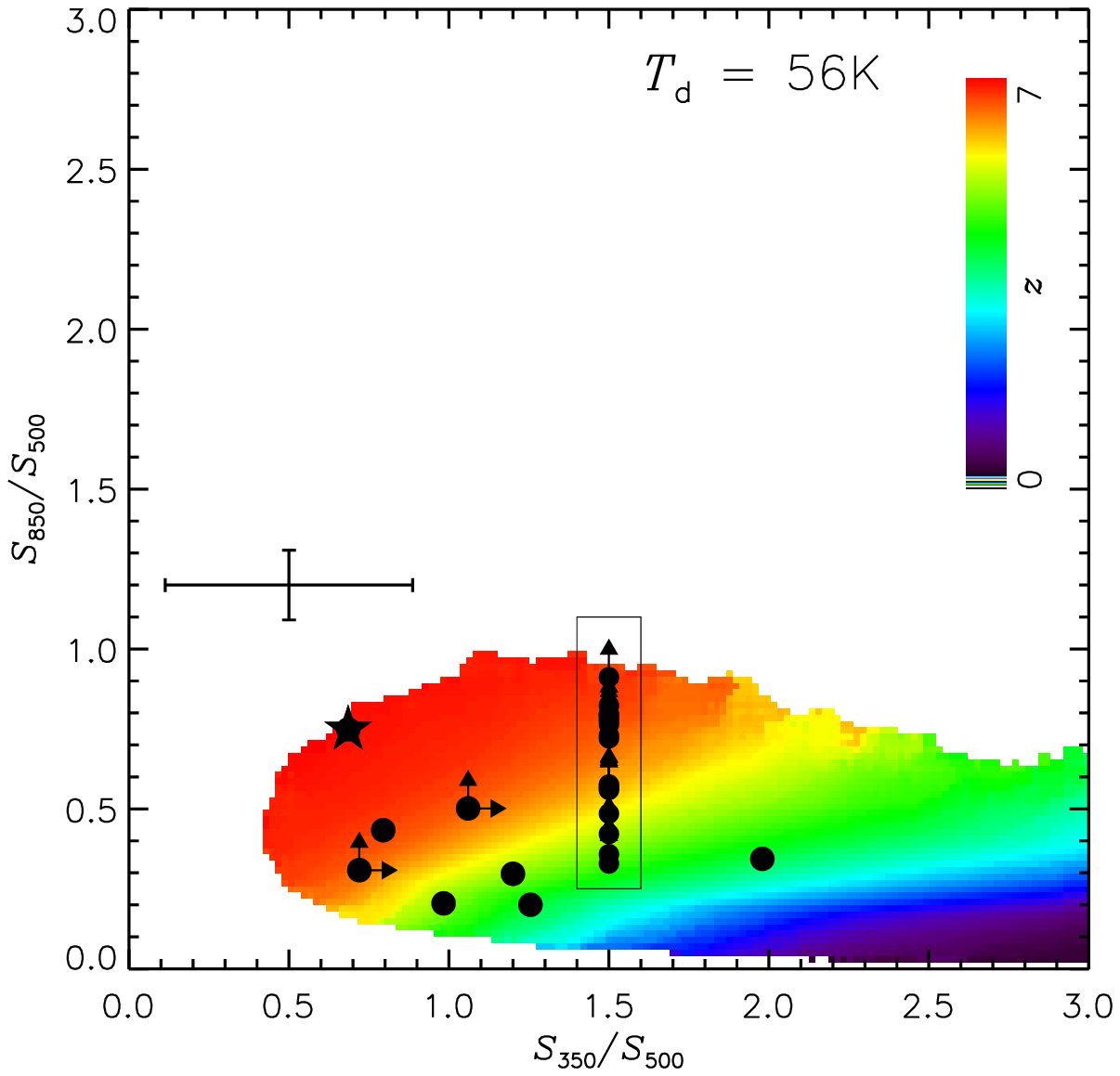,width=3.35in,angle=0}}
\vspace{3mm}
\centerline{\psfig{file=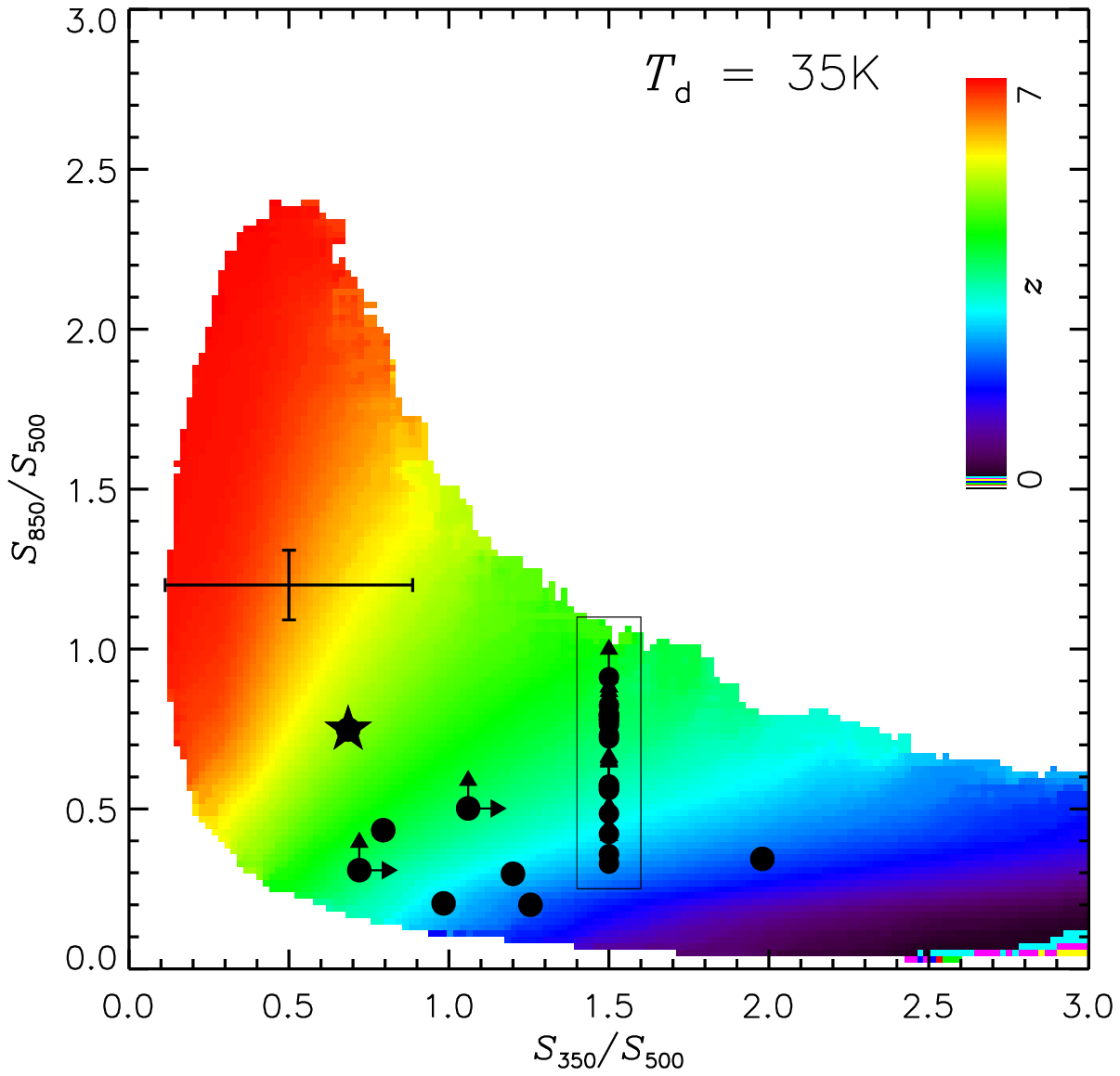,width=3.35in,angle=0}}
\vspace{-3mm} 
\noindent{\small\addtolength{\baselineskip}{-3pt}} 
\caption{Submm color-color plots for HFLS3 and its neighbouring SMGs,
  adapted from \citet{ivison12} and \citet{swinbank14}. For sources
  with flux densities below twice the confusion noise,
  \citep[$\sigma_{\rm conf}$,][]{nguyen10}, we show limits based on
  the measured flux density (zero, if negative) plus $\sigma_{\rm
    conf}$. The colored backgrounds indicate the typical redshift of
  the subset of $10^7$ model SEDs that fall in each pixel, where in
  the {\it top} panel we have adopted the dust temperature of HFLS3
  (\Td\ = 56\,K), a flat redshift distribution from $z=0$--7, a flat
  $\beta$ distribution from 1.5--2.0 (for HFLS3, $\beta=1.92\pm
  0.12$), and 10\% flux density errors. HFLS3 is the star symbol. The
  error bar is typical for sources detected at 350, 500 and
  850\,$\mu$m. Those sources without detections at 350 and 500\,$\mu$m
  are plotted arbitrarily at $S_{350}/S_{500} = 1.5$, inside a
  rectangular box; some could be considerably redder than this. Of
  those detected in both 350 and 500\,$\mu$m,
  S2FLS\,J170621.93+584826.8 is the reddest in $S_{350}/S_{500}$. A
  handful are as red as HFLS3 in $S_{850}/S_{500}$, e.g.\
  S2FLS\,J170647.80+584735.0, which is a 5.3-$\sigma$ SCUBA-2 source
  with no SPIRE counterpart. Despite its relatively high \Td, HFLS3 is
  the reddest of the sources detected in the three bands, though we
  note that redder objects may exist amongst the SPIRE-undetected
  SMGs. The {\it lower} plot shows a flat $\beta$ distribution from
  1.5--2.0 and \Td\ = 35\,K, illustrating how \Td\ influences the
  colors of an SMG.  Comparing the redshift constraints from both
  panels makes it clear that nothing can be usefully concluded about
  the likely redshifts from the far-infrared colors unless there are
  independent constraints on the dust temperature.}
\label{fig:cc}
\end{figure} 
 
The $K$-correction in the submm waveband means our SCUBA-2 maps are
sensitive to SMGs across a very wide redshift range, reducing the
contrast of any potential structure around HFLS3.  However, with
SCUBA-2, SPIRE and PACS photometry in hand, we are able to crudely
constrain the likely redshifts of the galaxies detected in the field
surrounding HFLS3, using their far-infrared/submm colors.
Fig.~\ref{fig:cc} shows color--color plots for HFLS3 and its
neighbouring SMGs, designed to exploit information from SCUBA-2 at
850\,$\mu$m to constrain the redshifts of galaxies at $z>2$
\citep{ivison12}, probing their colors across the rest-frame
$\approx$\,100-$\mu$m bump seen in the SEDs of all dusty, star-forming
galaxies. The colored backgrounds in the upper and lower panels of
Fig.~\ref{fig:cc} indicate the typical redshift of the subset of
$10^7$ model SEDs that fall in each pixel, where we have adopted a
flat redshift distribution ($z=0$--7), a flat distribution for the
spectral dependence of the dust emissivity ($\beta=1.8$--2.0, centered
on the $\beta$ measured for HFLS3) and 10\% flux density
uncertainties.  For the upper panel of Fig.~\ref{fig:cc} we adopt the
dust temperature of HFLS3 (\Td\ = 56\,K).

We concentrate only on those galaxies detected by SCUBA-2, since
SPIRE-detected galaxy with a typical SED in the vicinity of HFLS3
could not have evaded detection at 850\,$\mu$m. Despite its relatively
high \Td, HFLS3 is the reddest source detected in the three bands used
to make Fig.~\ref{fig:cc}.

For SMGs with SPIRE flux densities below $2\sigma_{\rm conf}$, we plot
limits based on the measured flux density (zero, if negative) plus
$\sigma_{\rm conf}$. We have placed those sources without detections
at 350 and 500\,$\mu$m at $S_{350}/S_{500} = 2$, arbitrarily; some
could be considerably redder than this in $S_{350}/S_{500}$, but we
cannot constrain this color reliably with the relatively shallow {\it
  Herschel} data at our disposal. Several SMGs may also be as red as
HFLS3 in $S_{850}/S_{500}$, perhaps redder.  One particularly
interesting example is S2FLS850\,J170647.80+584735.0, a 5.3-$\sigma$
SCUBA-2 source with no significant SPIRE emission. With
$S_{850}/S_{500}>0.9$, this SMG likely lies at $z>5$, with a lower
\Td\ and luminosity than HFLS3.

The lower panel of Fig.~\ref{fig:cc} shows the effect of lowering \Td,
illustrating an issue long-known to hamper studies of this kind:
far-infrared/submm colors are sensitive only to $(1+z)/T_{\rm d}$
\citep{blain99}, i.e.\ redshift and \Td\ are degenerate. As a result,
our current data does not allow us to conclude with certainty that the
environment surrounding HFLS3 contains other luminous, dusty
starbursts; neither can we rule it out.

\begin{figure}
\centerline{\psfig{file=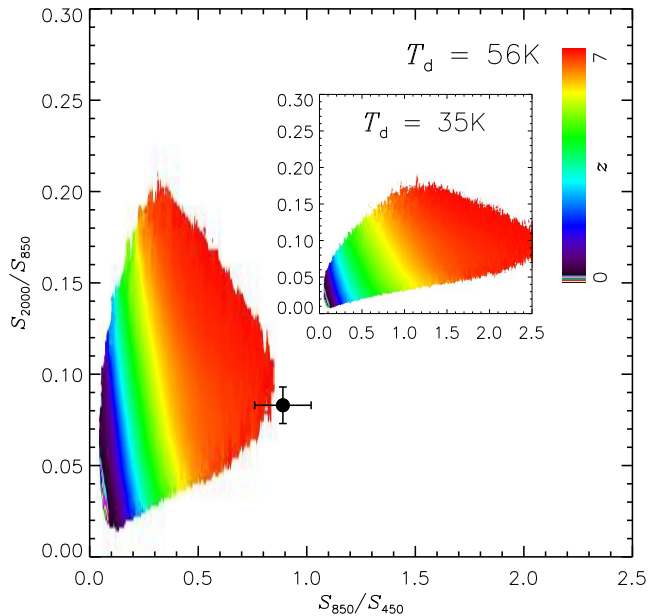,width=3.35in,angle=0}}
\vspace{-3mm} 
\noindent{\small\addtolength{\baselineskip}{-3pt}} 
\caption{$S_{850}/S_{450}$ versus $S_{2000}/S_{850}$, ratios for which
  deep, relatively unconfused and unbiased data can be obtained from
  the ground, covering tens of square arcminutes. The observed colors
  of HFLS3 are shown. The colored background indicates the typical
  redshift of the subset of $10^7$ model SEDs that fall in each pixel,
  where we have adopted the same dust parameters as
  Fig.~\ref{fig:cc}. {\it Inset:} Same plot, having changed only \Td,
  to 35\,K, illustrating the degeneracy between redshift and \Td.
  $S_{850}/S_{450}\ls 0.3$ suggests $z\ls 2$ for all but the warmest
  dust; $S_{850}/S_{450}\gs 1$ suggests $z\gs 5$, with dust much
  cooler than 35\,K unlikely at these redshifts; $S_{2000}/S_{850}$
  offers less insight.}
\label{fig:2mm}
\end{figure}

Single-dish imaging of this field at 450, 850, 1100 and 2000\,$\mu$m
is possible from the ground, reaching $\sigma_{450}\sim 2.5$\,mJy and
$\sigma_{2000}\sim 0.1$\,mJy over tens of arcmin$^2$ with existing
facilities in a few tens of hours. Would these data be capable of
further constraining the redshifts of the SMGs discovered here?
Fig.~\ref{fig:2mm} shows $S_{850}/S_{450}$ versus $S_{2000}/S_{850}$
and we see that the latter color offers little insight. For
$S_{850}/S_{450}\ls 0.3$ we can rule out $z\gs 2$ for all but the
warmest dust; $S_{850}/S_{450}\gs 1$ suggests $z\gs 5$, even for \Td\
= 35\,K, with significantly cooler dust unlikely in this redshift
regime. Deeper 450-$\mu$m imaging would therefore be useful.

%
% Figure 3
%

\subsection{Predictions from the Millennium Simulation}
\label{predictions}

Is the field surrounding HFLS3 less over-dense in submm sources than
expected for such a massive galaxy living in a biased environment at
high redshift, similar to those found around high-redshift radio
galaxies and radio-loud quasars at $z = 2$--4
\citep[e.g.][]{stevens03, stevens04}?  The answer to this question may
have implications for the potential gravitational amplification suffered
by HFLS3 (see \S1) or for investigating the potential presence of a
buried AGN and its role in supporting its high IR luminosity.

By necessity this comparison will be crude.  We therefore selected the
implementation of the \citet{bower06} galaxy-formation recipe in the
Millennium Simulation \citep{springel05} and searched the $z=6.2$
output for galaxies with a total baryonic mass in excess of $1.3\times
10^{11}$\,M$_{\odot}$, consistent with the combined mass of gas and
stars estimated for HFLS3 \citep{riechers13}.  By adopting a total
baryonic mass cut we are less sensitive to details of the early
star-formation histories of galaxies in the model.

We find just one galaxy in the $3.2\times 10^8$\,Mpc$^3$ volume at
$z=6.2$ with a total baryonic mass above $1.3\times
10^{11}$\,M$_{\odot}$. All its baryonic mass is in stars; it hosts a
$2\times 10^8$-M$_{\odot}$ black hole and is the central galaxy of a
$6\times 10^{12}$-M$_{\odot}$ halo, $4\times$ more massive than the
next most massive galaxy's halo within the volume, and the optimal
environment to find merging galaxies according to the simulations of
\citet{hopkins08a}. Another 16 galaxies are spread across a
$\sim$\,0.7-comoving-Mpc-diameter region around the most massive
galaxy, but most of these are dwarf galaxies with baryonic masses,
$\ls 10^9$\,M$_\odot$. Inside a sphere with an angular size of 9$'$
diameter, centered on the $1.3\times 10^{11}$-M$_{\odot}$ galaxy, only
two galaxies have total baryonic masses of $\gs$\,15\% of the mass of
HFLS3 (we choose this limit as the faintest submm emitters in this
field have 850-$\mu$m flux densities of around 15\% that of HFLS3);
the next most massive galaxy has half this mass. The total masses of
these two companion galaxies in stars and gas are approximately 3 and
$4 \times 10^{10}$\,M$_{\odot}$ and their predicted $K_{\rm Vega}$
magnitudes are 23.5 and 25.0. The HFLS3 clone is predicted to have
$K_{\rm Vega}=22.5$ for a distance modulus of 49.0, about a magnitude
fainter than observed.

From this simple theoretical comparison, we thus expect $2\pm 2$
detectable galaxies in the vicinity of HFLS3 if, as expected, their
starburst lifetimes are a significant fraction of the time available
at this early epoch. This is consistent with the fact that some
high-redshift SMGs do have submm-bright companions \citep[e.g.\ GN\,20
--][]{daddi09} while others have Lyman-break galaxies nearby but no
submm-bright companions \citep[e.g.\ AzTEC-3 --][]{capak11}.

Having found no clear evidence for or against the level of
over-density expected in simulations, we can draw no strong
conclusions regarding the likely gravitational amplification suffered
by HFLS3, or for the likely fraction of its luminosity provided by a
buried AGN.

\section{Conclusions}
\label{conclusions}
 
We have detected the most distant, dusty starburst galaxy, HFLS3, at
high significance with SCUBA-2. We detect another 29 dusty galaxies
within an area of 67.2\,arcmin$^2$ surrounding HFLS3, most of them
likely at lower redshift.  We find no compelling evidence, from
surface density or color, for an over-density of SMGs around HFLS3,
although applying similar selection criteria to theoretical models
suggests that a modest excess could be expected, as is found for some
other high-redshift SMGs \citep[e.g.\ GN\,20 --][]{daddi09}. We can
therefore draw no strong conclusions regarding the likely
gravitational amplification suffered by HFLS3, or for the likely
fraction of its luminosity provided by a buried AGN.

\section*{Acknowledgements}

We thank John Helly for help with the Millennium Simulation.  RJI and
IRS acknowledge support from the European Research Council (ERC) in
the form of Advanced Investigator programs, {\sc cosmicism} and {\sc
  dustygal}, respectively.  IRS also acknowledges support from the
UK's Science and Technology Facilities Council (STFC, ST/I001573/1), a
Leverhulme Fellowship and a Royal Society/Wolfson Merit Award. JEG
acknowledges the Royal Society for support. The
Dark Cosmology Centre is funded by the Danish National Research
Foundation. The JCMT is operated by the Joint Astronomy Centre on
behalf of STFC, the National Research Council of Canada and (until 31
March 2013) the Netherlands Organisation for Scientific
Research. Additional funds for the construction of SCUBA-2 were
provided by the Canada Foundation for Innovation.  {\it Herschel} is
an ESA space observatory with science instruments provided by
European-led Principal Investigator consortia and with important
participation from NASA. SPIRE was developed by a consortium of
institutes led by Cardiff Univ.\ (UK) and including: Univ.\ Lethbridge
(Canada); NAOC (China); CEA, LAM (France); IFSI, Univ.\ Padua (Italy);
IAC (Spain); Stockholm Observatory (Sweden); Imperial College London,
RAL, UCL-MSSL, UKATC, Univ.\ Sussex (UK); and Caltech, JPL, NHSC,
Univ. Colorado (USA). This development has been supported by national
funding agencies: CSA (Canada); NAOC (China); CEA, CNES, CNRS
(France); ASI (Italy); MCINN (Spain); SNSB (Sweden); STFC, UKSA (UK);
and NASA (USA).  The JCMT and {\it Herschel} data used in this paper
can be obtained from the JCMT archive
({www.jach.hawaii.edu/JCMT/archive}) and the {\em Herschel} Database
in Marseille, HeDaM, ({hedam.oamp.fr/HerMES}), respectively.

{\it Facilities:} \facility{JCMT, {\it Herschel}}

\bibliographystyle{apj}
\bibliography{robson}

\begin{thebibliography}{39}
\expandafter\ifx\csname natexlab\endcsname\relax\def\natexlab#1{#1}\fi

\bibitem[{{Blain}(1999)}]{blain99}
{Blain}, A.~W. 1999, \mnras, 309, 955

\bibitem[{{Blain} \& {Longair}(1993)}]{blainlongair93}
{Blain}, A.~W., \& {Longair}, M.~S. 1993, \mnras, 264, 509

\bibitem[{{Bower} {et~al.}(2006){Bower}, {Benson}, {Malbon}, {Helly}, {Frenk},
  {Baugh}, {Cole}, \& {Lacey}}]{bower06}
{Bower}, R.~G., {Benson}, A.~J., {Malbon}, R., {et~al.} 2006, \mnras, 370, 645

\bibitem[{{Capak} {et~al.}(2011){Capak}, {Riechers}, {Scoville}, {Carilli},
  {Cox}, {Neri}, {Robertson}, {Salvato}, {Schinnerer}, {Yan}, {Wilson}, {Yun},
  {Civano}, {Elvis}, {Karim}, {Mobasher}, \& {Staguhn}}]{capak11}
{Capak}, P.~L., {Riechers}, D., {Scoville}, N.~Z., {et~al.} 2011, \nat, 470,
  233

\bibitem[{{Cavanagh} {et~al.}(2008){Cavanagh}, {Jenness}, {Economou}, \&
  {Currie}}]{cavanagh08}
{Cavanagh}, B., {Jenness}, T., {Economou}, F., \& {Currie}, M.~J. 2008,
  Astronomische Nachrichten, 329, 295

\bibitem[{{Chabrier}(2003)}]{chabrier03}
{Chabrier}, G. 2003, \pasp, 115, 763

\bibitem[{{Chapin} {et~al.}(2013){Chapin}, {Berry}, {Gibb}, {Jenness}, {Scott},
  {Tilanus}, {Economou}, \& {Holland}}]{chapin13}
{Chapin}, E.~L., {Berry}, D.~S., {Gibb}, A.~G., {et~al.} 2013, \mnras, 430,
  2545

\bibitem[{{Chapin} {et~al.}(2011){Chapin}, {Chapman}, {Coppin}, {Devlin},
  {Dunlop}, {Greve}, {Halpern}, {Hasselfield}, {Hughes}, {Ivison}, {Marsden},
  {Moncelsi}, {Netterfield}, {Pascale}, {Scott}, {Smail}, {Viero}, {Walter},
  {Weiss}, \& {van der Werf}}]{chapin11}
{Chapin}, E.~L., {Chapman}, S.~C., {Coppin}, K.~E., {et~al.} 2011, \mnras, 411,
  505

\bibitem[{{Combes} {et~al.}(2012){Combes}, {Rex}, {Rawle}, {Egami}, {Boone},
  {Smail}, {Richard}, {Ivison}, {Gurwell}, {Casey}, {Omont}, {Berciano Alba},
  {Dessauges-Zavadsky}, {Edge}, {Fazio}, {Kneib}, {Okabe}, {Pell{\'o}},
  {P{\'e}rez-Gonz{\'a}lez}, {Schaerer}, {Smith}, {Swinbank}, \& {van der
  Werf}}]{combes12}
{Combes}, F., {Rex}, M., {Rawle}, T.~D., {et~al.} 2012, \aap, 538, L4

\bibitem[{{Cooray} {et~al.}(2014){Cooray}, {Calanog}, {Wardlow}, {Bock},
  {Bridge}, {Burgarella}, {Bussmann}, {Casey}, {Clements}, {Conley}, {Farrah},
  {Fu}, {Gavazzi}, {Ivison}, {Laporte}, {Lofaro}, {Ma}, {Magdis}, {Oliver},
  {Osage}, {Pe'rez-Fournon}, {Riechers}, {Rigopoulou}, {Scott}, {Viero}, \&
  {Watson}}]{cooray14}
{Cooray}, A., {Calanog}, J., {Wardlow}, J.~L., {et~al.} 2014, ArXiv e-prints,
  arXiv:1404.1378

\bibitem[{{Coppin} {et~al.}(2006){Coppin}, {Chapin}, {Mortier}, {Scott},
  {Borys}, {Dunlop}, {Halpern}, {Hughes}, {Pope}, {Scott}, {Serjeant}, {Wagg},
  {Alexander}, {Almaini}, {Aretxaga}, {Babbedge}, {Best}, {Blain}, {Chapman},
  {Clements}, {Crawford}, {Dunne}, {Eales}, {Edge}, {Farrah}, {Gazta{\~n}aga},
  {Gear}, {Granato}, {Greve}, {Fox}, {Ivison}, {Jarvis}, {Jenness}, {Lacey},
  {Lepage}, {Mann}, {Marsden}, {Martinez-Sansigre}, {Oliver}, {Page},
  {Peacock}, {Pearson}, {Percival}, {Priddey}, {Rawlings}, {Rowan-Robinson},
  {Savage}, {Seigar}, {Sekiguchi}, {Silva}, {Simpson}, {Smail}, {Stevens},
  {Takagi}, {Vaccari}, {van Kampen}, \& {Willott}}]{coppin06}
{Coppin}, K., {Chapin}, E.~L., {Mortier}, A.~M.~J., {et~al.} 2006, \mnras, 372,
  1621

\bibitem[{{Cox} {et~al.}(2011){Cox}, {Krips}, {Neri}, {Omont}, {G{\"u}sten},
  {Menten}, {Wyrowski}, {Wei{\ss}}, {Beelen}, {Gurwell}, {Dannerbauer},
  {Ivison}, {Negrello}, {Aretxaga}, {Hughes}, {Auld}, {Baes}, {Blundell},
  {Buttiglione}, {Cava}, {Cooray}, {Dariush}, {Dunne}, {Dye}, {Eales},
  {Frayer}, {Fritz}, {Gavazzi}, {Hopwood}, {Ibar}, {Jarvis}, {Maddox},
  {Micha{l}owski}, {Pascale}, {Pohlen}, {Rigby}, {Smith}, {Swinbank}, {Temi},
  {Valtchanov}, {van der Werf}, \& {de Zotti}}]{cox11}
{Cox}, P., {Krips}, M., {Neri}, R., {et~al.} 2011, \apj, 740, 63

\bibitem[{{Daddi} {et~al.}(2009){Daddi}, {Dannerbauer}, {Stern}, {Dickinson},
  {Morrison}, {Elbaz}, {Giavalisco}, {Mancini}, {Pope}, \& {Spinrad}}]{daddi09}
{Daddi}, E., {Dannerbauer}, H., {Stern}, D., {et~al.} 2009, \apj, 694, 1517

\bibitem[{{Dempsey} {et~al.}(2013){Dempsey}, {Friberg}, {Jenness}, {Tilanus},
  {Thomas}, {Holland}, {Bintley}, {Berry}, {Chapin}, {Chrysostomou}, {Davis},
  {Gibb}, {Parsons}, \& {Robson}}]{dempsey13}
{Dempsey}, J.~T., {Friberg}, P., {Jenness}, T., {et~al.} 2013, \mnras, 430,
  2534

\bibitem[{{Dowell} {et~al.}(2014){Dowell}, {Conley}, {Glenn}, {Arumugam},
  {Asboth}, {Aussel}, {Bertoldi}, {B{\'e}thermin}, {Bock}, {Boselli}, {Bridge},
  {Buat}, {Burgarella}, {Cabrera-Lavers}, {Casey}, {Chapman}, {Clements},
  {Conversi}, {Cooray}, {Dannerbauer}, {De Bernardis}, {Ellsworth-Bowers},
  {Farrah}, {Franceschini}, {Griffin}, {Gurwell}, {Halpern}, {Hatziminaoglou},
  {Heinis}, {Ibar}, {Ivison}, {Laporte}, {Marchetti},
  {Mart{\'{\i}}nez-Navajas}, {Marsden}, {Morrison}, {Nguyen}, {O'Halloran},
  {Oliver}, {Omont}, {Page}, {Papageorgiou}, {Pearson}, {Petitpas},
  {P{\'e}rez-Fournon}, {Pohlen}, {Riechers}, {Rigopoulou}, {Roseboom},
  {Rowan-Robinson}, {Sayers}, {Schulz}, {Scott}, {Seymour}, {Shupe}, {Smith},
  {Streblyanska}, {Symeonidis}, {Vaccari}, {Valtchanov}, {Vieira}, {Viero},
  {Wang}, {Wardlow}, {Xu}, \& {Zemcov}}]{dowell14}
{Dowell}, C.~D., {Conley}, A., {Glenn}, J., {et~al.} 2014, \apj, 780, 75

\bibitem[{{Franceschini} {et~al.}(1991){Franceschini}, {Toffolatti}, {Mazzei},
  {Danese}, \& {de Zotti}}]{fran91}
{Franceschini}, A., {Toffolatti}, L., {Mazzei}, P., {Danese}, L., \& {de
  Zotti}, G. 1991, \aaps, 89, 285

\bibitem[{{Geach} {et~al.}(2013){Geach}, {Chapin}, {Coppin}, {Dunlop},
  {Halpern}, {Smail}, {Werf}, {Serjeant}, {Farrah}, {Roseboom}, {Targett},
  {Arumugam}, {Asboth}, {Blain}, {Chrysostomou}, {Clarke}, {Ivison}, {Jones},
  {Karim}, {Mackenzie}, {Meijerink}, {Michalowski}, {Scott}, {Simpson},
  {Swinbank}, {Alexander}, {Almaini}, {Aretxaga}, {Best}, {Chapman},
  {Clements}, {Conselice}, {Danielson}, {Eales}, {Edge}, {Gibb}, {Hughes},
  {Jenness}, {Knudsen}, {Lacey}, {Marsden}, {McMahon}, {Oliver}, {Page},
  {Peacock}, {Rigopoulou}, {Robson}, {Spaans}, {Stevens}, {Webb}, {Willott},
  {Wilson}, \& {Zemcov}}]{geach13}
{Geach}, J.~E., {Chapin}, E.~L., {Coppin}, K.~E.~K., {et~al.} 2013, \mnras,
  432, 53

\bibitem[{{Griffin} {et~al.}(2010){Griffin}, {Abergel}, {Abreu}, {Ade},
  {Andr{\'e}}, {Augueres}, {Babbedge}, {Bae}, {Baillie}, {Baluteau}, {Barlow},
  {Bendo}, {Benielli}, {Bock}, {Bonhomme}, {Brisbin}, {Brockley-Blatt},
  {Caldwell}, {Cara}, {Castro-Rodriguez}, {Cerulli}, {Chanial}, {Chen},
  {Clark}, {Clements}, {Clerc}, {Coker}, {Communal}, {Conversi}, {Cox},
  {Crumb}, {Cunningham}, {Daly}, {Davis}, {de Antoni}, {Delderfield}, {Devin},
  {di Giorgio}, {Didschuns}, {Dohlen}, {Donati}, {Dowell}, {Dowell}, {Duband},
  {Dumaye}, {Emery}, {Ferlet}, {Ferrand}, {Fontignie}, {Fox}, {Franceschini},
  {Frerking}, {Fulton}, {Garcia}, {Gastaud}, {Gear}, {Glenn}, {Goizel},
  {Griffin}, {Grundy}, {Guest}, {Guillemet}, {Hargrave}, {Harwit}, {Hastings},
  {Hatziminaoglou}, {Herman}, {Hinde}, {Hristov}, {Huang}, {Imhof}, {Isaak},
  {Israelsson}, {Ivison}, {Jennings}, {Kiernan}, {King}, {Lange}, {Latter},
  {Laurent}, {Laurent}, {Leeks}, {Lellouch}, {Levenson}, {Li}, {Li},
  {Lilienthal}, {Lim}, {Liu}, {Lu}, {Madden}, {Mainetti}, {Marliani}, {McKay},
  {Mercier}, {Molinari}, {Morris}, {Moseley}, {Mulder}, {Mur}, {Naylor},
  {Nguyen}, {O'Halloran}, {Oliver}, {Olofsson}, {Olofsson}, {Orfei}, {Page},
  {Pain}, {Panuzzo}, {Papageorgiou}, {Parks}, {Parr-Burman}, {Pearce},
  {Pearson}, {P{\'e}rez-Fournon}, {Pinsard}, {Pisano}, {Podosek}, {Pohlen},
  {Polehampton}, {Pouliquen}, {Rigopoulou}, {Rizzo}, {Roseboom}, {Roussel},
  {Rowan-Robinson}, {Rownd}, {Saraceno}, {Sauvage}, {Savage}, {Savini},
  {Sawyer}, {Scharmberg}, {Schmitt}, {Schneider}, {Schulz}, {Schwartz},
  {Shafer}, {Shupe}, {Sibthorpe}, {Sidher}, {Smith}, {Smith}, {Smith},
  {Spencer}, {Stobie}, {Sudiwala}, {Sukhatme}, {Surace}, {Stevens}, {Swinyard},
  {Trichas}, {Tourette}, {Triou}, {Tseng}, {Tucker}, {Turner}, {Vaccari},
  {Valtchanov}, {Vigroux}, {Virique}, {Voellmer}, {Walker}, {Ward}, {Waskett},
  {Weilert}, {Wesson}, {White}, {Whitehouse}, {Wilson}, {Winter}, {Woodcraft},
  {Wright}, {Xu}, {Zavagno}, {Zemcov}, {Zhang}, \& {Zonca}}]{griffin10}
{Griffin}, M.~J., {Abergel}, A., {Abreu}, A., {et~al.} 2010, \aap, 518, L3+

\bibitem[{{Holland} {et~al.}(1999){Holland}, {Robson}, {Gear}, {Cunningham},
  {Lightfoot}, {Jenness}, {Ivison}, {Stevens}, {Ade}, {Griffin}, {Duncan},
  {Murphy}, \& {Naylor}}]{holland99}
{Holland}, W.~S., {Robson}, E.~I., {Gear}, W.~K., {et~al.} 1999, \mnras, 303,
  659

\bibitem[{{Holland} {et~al.}(2013){Holland}, {Bintley}, {Chapin},
  {Chrysostomou}, {Davis}, {Dempsey}, {Duncan}, {Fich}, {Friberg}, {Halpern},
  {Irwin}, {Jenness}, {Kelly}, {MacIntosh}, {Robson}, {Scott}, {Ade},
  {Atad-Ettedgui}, {Berry}, {Craig}, {Gao}, {Gibb}, {Hilton}, {Hollister},
  {Kycia}, {Lunney}, {McGregor}, {Montgomery}, {Parkes}, {Tilanus}, {Ullom},
  {Walther}, {Walton}, {Woodcraft}, {Amiri}, {Atkinson}, {Burger}, {Chuter},
  {Coulson}, {Doriese}, {Dunare}, {Economou}, {Niemack}, {Parsons},
  {Reintsema}, {Sibthorpe}, {Smail}, {Sudiwala}, \& {Thomas}}]{holland13}
{Holland}, W.~S., {Bintley}, D., {Chapin}, E.~L., {et~al.} 2013, \mnras, 430,
  2513

\bibitem[{{Hopkins} {et~al.}(2008){Hopkins}, {Hernquist}, {Cox}, \& {Kere{\v
  s}}}]{hopkins08a}
{Hopkins}, P.~F., {Hernquist}, L., {Cox}, T.~J., \& {Kere{\v s}}, D. 2008,
  \apjs, 175, 356

\bibitem[{{Ivison} {et~al.}(2000){Ivison}, {Dunlop}, {Smail}, {Dey}, {Liu}, \&
  {Graham}}]{ivison00}
{Ivison}, R.~J., {Dunlop}, J.~S., {Smail}, I., {et~al.} 2000, \apj, 542, 27

\bibitem[{{Ivison} {et~al.}(2012){Ivison}, {Smail}, {Amblard}, {Arumugam}, {De
  Breuck}, {Emonts}, {Feain}, {Greve}, {Haas}, {Ibar}, {Jarvis}, {Kov{\'a}cs},
  {Lehnert}, {Nesvadba}, {R{\"o}ttgering}, {Seymour}, \&
  {Wylezalek}}]{ivison12}
{Ivison}, R.~J., {Smail}, I., {Amblard}, A., {et~al.} 2012, \mnras, 425, 1320

\bibitem[{{Jenness} {et~al.}(2010){Jenness}, {Robson}, \&
  {Stevens}}]{jenness10}
{Jenness}, T., {Robson}, E.~I., \& {Stevens}, J.~A. 2010, \mnras, 401, 1240

\bibitem[{{Kaufman} {et~al.}(1999){Kaufman}, {Wolfire}, {Hollenbach}, \&
  {Luhman}}]{kaufman99}
{Kaufman}, M.~J., {Wolfire}, M.~G., {Hollenbach}, D.~J., \& {Luhman}, M.~L.
  1999, \apj, 527, 795

\bibitem[{{Nguyen} {et~al.}(2010){Nguyen}, {Schulz}, {Levenson}, {Amblard},
  {Arumugam}, {Aussel}, {Babbedge}, {Blain}, {Bock}, {Boselli}, {Buat},
  {Castro-Rodriguez}, {Cava}, {Chanial}, {Chapin}, {Clements}, {Conley},
  {Conversi}, {Cooray}, {Dowell}, {Dwek}, {Eales}, {Elbaz}, {Fox},
  {Franceschini}, {Gear}, {Glenn}, {Griffin}, {Halpern}, {Hatziminaoglou},
  {Ibar}, {Isaak}, {Ivison}, {Lagache}, {Lu}, {Madden}, {Maffei}, {Mainetti},
  {Marchetti}, {Marsden}, {Marshall}, {O'Halloran}, {Oliver}, {Omont}, {Page},
  {Panuzzo}, {Papageorgiou}, {Pearson}, {Perez Fournon}, {Pohlen}, {Rangwala},
  {Rigopoulou}, {Rizzo}, {Roseboom}, {Rowan-Robinson}, {Scott}, {Seymour},
  {Shupe}, {Smith}, {Stevens}, {Symeonidis}, {Trichas}, {Tugwell}, {Vaccari},
  {Valtchanov}, {Vigroux}, {Wang}, {Ward}, {Wiebe}, {Wright}, {Xu}, \&
  {Zemcov}}]{nguyen10}
{Nguyen}, H.~T., {Schulz}, B., {Levenson}, L., {et~al.} 2010, \aap, 518, L5+

\bibitem[{{Oliver} {et~al.}(2012){Oliver}, {Bock}, {Altieri}, {Amblard},
  {Arumugam}, {Aussel}, {Babbedge}, {Beelen}, {B{\'e}thermin}, {Blain},
  {Boselli}, {Bridge}, {Brisbin}, {Buat}, {Burgarella},
  {Castro-Rodr{\'{\i}}guez}, {Cava}, {Chanial}, {Cirasuolo}, {Clements},
  {Conley}, {Conversi}, {Cooray}, {Dowell}, {Dubois}, {Dwek}, {Dye}, {Eales},
  {Elbaz}, {Farrah}, {Feltre}, {Ferrero}, {Fiolet}, {Fox}, {Franceschini},
  {Gear}, {Giovannoli}, {Glenn}, {Gong}, {Gonz{\'a}lez Solares}, {Griffin},
  {Halpern}, {Harwit}, {Hatziminaoglou}, {Heinis}, {Hurley}, {Hwang}, {Hyde},
  {Ibar}, {Ilbert}, {Isaak}, {Ivison}, {Lagache}, {Le Floc'h}, {Levenson},
  {Faro}, {Lu}, {Madden}, {Maffei}, {Magdis}, {Mainetti}, {Marchetti},
  {Marsden}, {Marshall}, {Mortier}, {Nguyen}, {O'Halloran}, {Omont}, {Page},
  {Panuzzo}, {Papageorgiou}, {Patel}, {Pearson}, {P{\'e}rez-Fournon}, {Pohlen},
  {Rawlings}, {Raymond}, {Rigopoulou}, {Riguccini}, {Rizzo}, {Rodighiero},
  {Roseboom}, {Rowan-Robinson}, {S{\'a}nchez Portal}, {Schulz}, {Scott},
  {Seymour}, {Shupe}, {Smith}, {Stevens}, {Symeonidis}, {Trichas}, {Tugwell},
  {Vaccari}, {Valtchanov}, {Vieira}, {Viero}, {Vigroux}, {Wang}, {Ward},
  {Wardlow}, {Wright}, {Xu}, \& {Zemcov}}]{oliver12}
{Oliver}, S.~J., {Bock}, J., {Altieri}, B., {et~al.} 2012, \mnras, 424, 1614

\bibitem[{{Pilbratt} {et~al.}(2010){Pilbratt}, {Riedinger}, {Passvogel},
  {Crone}, {Doyle}, {Gageur}, {Heras}, {Jewell}, {Metcalfe}, {Ott}, \&
  {Schmidt}}]{pilbratt10}
{Pilbratt}, G.~L., {Riedinger}, J.~R., {Passvogel}, T., {et~al.} 2010, \aap,
  518, L1+

\bibitem[{{Poglitsch} {et~al.}(2010){Poglitsch}, {Waelkens}, {Geis},
  {Feuchtgruber}, {Vandenbussche}, {Rodriguez}, {Krause}, {Renotte}, {van
  Hoof}, {Saraceno}, {Cepa}, {Kerschbaum}, {Agn{\`e}se}, {Ali}, {Altieri},
  {Andreani}, {Augueres}, {Balog}, {Barl}, {Bauer}, {Belbachir}, {Benedettini},
  {Billot}, {Boulade}, {Bischof}, {Blommaert}, {Callut}, {Cara}, {Cerulli},
  {Cesarsky}, {Contursi}, {Creten}, {De Meester}, {Doublier}, {Doumayrou},
  {Duband}, {Exter}, {Genzel}, {Gillis}, {Gr{\"o}zinger}, {Henning},
  {Herreros}, {Huygen}, {Inguscio}, {Jakob}, {Jamar}, {Jean}, {de Jong},
  {Katterloher}, {Kiss}, {Klaas}, {Lemke}, {Lutz}, {Madden}, {Marquet},
  {Martignac}, {Mazy}, {Merken}, {Montfort}, {Morbidelli}, {M{\"u}ller},
  {Nielbock}, {Okumura}, {Orfei}, {Ottensamer}, {Pezzuto}, {Popesso},
  {Putzeys}, {Regibo}, {Reveret}, {Royer}, {Sauvage}, {Schreiber}, {Stegmaier},
  {Schmitt}, {Schubert}, {Sturm}, {Thiel}, {Tofani}, {Vavrek}, {Wetzstein},
  {Wieprecht}, \& {Wiezorrek}}]{poglitsch10}
{Poglitsch}, A., {Waelkens}, C., {Geis}, N., {et~al.} 2010, \aap, 518, L2+

\bibitem[{{Priddey} {et~al.}(2008){Priddey}, {Ivison}, \& {Isaak}}]{priddey08}
{Priddey}, R.~S., {Ivison}, R.~J., \& {Isaak}, K.~G. 2008, \mnras, 383, 289

\bibitem[{{Rawle} {et~al.}(2014){Rawle}, {Egami}, {Bussmann}, {Gurwell},
  {Ivison}, {Boone}, {Combes}, {Danielson}, {Rex}, {Richard}, {Smail},
  {Swinbank}, {Altieri}, {Blain}, {Clement}, {Dessauges-Zavadsky}, {Edge},
  {Fazio}, {Jones}, {Kneib}, {Omont}, {P{\'e}rez-Gonz{\'a}lez}, {Schaerer},
  {Valtchanov}, {van der Werf}, {Walth}, {Zamojski}, \& {Zemcov}}]{rawle14}
{Rawle}, T.~D., {Egami}, E., {Bussmann}, R.~S., {et~al.} 2014, \apj, 783, 59

\bibitem[{{Riechers} {et~al.}(2013){Riechers}, {Bradford}, {Clements},
  {Dowell}, {P{\'e}rez-Fournon}, {Ivison}, {Bridge}, {Conley}, {Fu}, {Vieira},
  {Wardlow}, {Calanog}, {Cooray}, {Hurley}, {Neri}, {Kamenetzky}, {Aguirre},
  {Altieri}, {Arumugam}, {Benford}, {B{\'e}thermin}, {Bock}, {Burgarella},
  {Cabrera-Lavers}, {Chapman}, {Cox}, {Dunlop}, {Earle}, {Farrah}, {Ferrero},
  {Franceschini}, {Gavazzi}, {Glenn}, {Solares}, {Gurwell}, {Halpern},
  {Hatziminaoglou}, {Hyde}, {Ibar}, {Kov{\'a}cs}, {Krips}, {Lupu}, {Maloney},
  {Martinez-Navajas}, {Matsuhara}, {Murphy}, {Naylor}, {Nguyen}, {Oliver},
  {Omont}, {Page}, {Petitpas}, {Rangwala}, {Roseboom}, {Scott}, {Smith},
  {Staguhn}, {Streblyanska}, {Thomson}, {Valtchanov}, {Viero}, {Wang},
  {Zemcov}, \& {Zmuidzinas}}]{riechers13}
{Riechers}, D.~A., {Bradford}, C.~M., {Clements}, D.~L., {et~al.} 2013, \nat,
  496, 329

\bibitem[{{Robson} {et~al.}(2004){Robson}, {Priddey}, {Isaak}, \&
  {McMahon}}]{robson04}
{Robson}, I., {Priddey}, R.~S., {Isaak}, K.~G., \& {McMahon}, R.~G. 2004,
  \mnras, 351, L29

\bibitem[{{Springel} {et~al.}(2005){Springel}, {White}, {Jenkins}, {Frenk},
  {Yoshida}, {Gao}, {Navarro}, {Thacker}, {Croton}, {Helly}, {Peacock}, {Cole},
  {Thomas}, {Couchman}, {Evrard}, {Colberg}, \& {Pearce}}]{springel05}
{Springel}, V., {White}, S.~D.~M., {Jenkins}, A., {et~al.} 2005, \nat, 435, 629

\bibitem[{{Stevens} {et~al.}(2010){Stevens}, {Jarvis}, {Coppin}, {Page},
  {Greve}, {Carrera}, \& {Ivison}}]{stevens10}
{Stevens}, J.~A., {Jarvis}, M.~J., {Coppin}, K.~E.~K., {et~al.} 2010, \mnras,
  405, 2623

\bibitem[{{Stevens} {et~al.}(2004){Stevens}, {Page}, {Ivison}, {Smail}, \&
  {Carrera}}]{stevens04}
{Stevens}, J.~A., {Page}, M.~J., {Ivison}, R.~J., {Smail}, I., \& {Carrera},
  F.~J. 2004, \apjl, 604, L17

\bibitem[{{Stevens} {et~al.}(2003){Stevens}, {Ivison}, {Dunlop}, {Smail},
  {Percival}, {Hughes}, {R{\"o}ttgering}, {van Breugel}, \&
  {Reuland}}]{stevens03}
{Stevens}, J.~A., {Ivison}, R.~J., {Dunlop}, J.~S., {et~al.} 2003, \nat, 425,
  264

\bibitem[{{Swinbank} {et~al.}(2014){Swinbank}, {Simpson}, {Smail}, {Harrison},
  {Hodge}, {Karim}, {Walter}, {Alexander}, {Brandt}, {de Breuck}, {da Cunha},
  {Chapman}, {Coppin}, {Danielson}, {Dannerbauer}, {Decarli}, {Greve},
  {Ivison}, {Knudsen}, {Lagos}, {Schinnerer}, {Thomson}, {Wardlow}, {Wei{\ss}},
  \& {van der Werf}}]{swinbank14}
{Swinbank}, A.~M., {Simpson}, J.~M., {Smail}, I., {et~al.} 2014, \mnras, 438,
  1267

\bibitem[{{Wei{\ss}} {et~al.}(2009){Wei{\ss}}, {Kov{\'a}cs}, {Coppin}, {Greve},
  {Walter}, {Smail}, {Dunlop}, {Knudsen}, {Alexander}, {Bertoldi}, {Brandt},
  {Chapman}, {Cox}, {Dannerbauer}, {De Breuck}, {Gawiser}, {Ivison}, {Lutz},
  {Menten}, {Koekemoer}, {Kreysa}, {Kurczynski}, {Rix}, {Schinnerer}, \& {van
  der Werf}}]{weiss09}
{Wei{\ss}}, A., {Kov{\'a}cs}, A., {Coppin}, K., {et~al.} 2009, \apj, 707, 1201

\end{thebibliography}

\end{document}